\documentclass[aps,prl,twocolumn,groupedaddress]{revtex4}
\usepackage{graphicx}

\begin{document}


\title{Voltage Dependence of Spin Polarized Tunneling}


\author{S.O. Valenzuela, D.J. Monsma, C.M. Marcus, V. Narayanamurti and M. Tinkham}
\affiliation{Gordon McKay Laboratory, Harvard University,
Cambridge, Massachusetts 02138}

\author{}
\affiliation{}


\date{\today}

\begin{abstract}

A mesoscopic spin valve is used to determine the effective spin
polarization of electrons tunneling from and into ferromagnetic
transition metals at finite voltages. The tunneling spin
polarization \emph{from} the ferromagnet (FM) slowly decreases
with bias, but drops faster and even inverts with voltage when
electrons tunnel \emph{into} it. A bias-dependent free electron
model shows that in the former case electrons originate near the
Fermi level of the FM with large polarization whereas in the
latter, electrons tunnel into hot electron states for which the
polarization is significantly reduced. The change in sign is
ascribed to the detailed matching of the electron wave function
through the tunnel barrier.

\end{abstract}


\maketitle


Magnetic tunnel junctions (MTJs) \cite{Jul,moo1,miya} typically
consist of ferromagnet-insulator-ferromagnet (FM-I-FM) structures,
often in the form CoFe-Al$_{2}$O$_{3}$-NiFe. The tunneling of
electrons across the insulating barrier is spin polarized, and
causes the so-called tunnel magnetoresistance (TMR). As an
important group of ``spintronics" devices, magnetic tunnel
junctions are already finding applications in magnetic field
sensors and non-volatile magnetic random access memories (MRAM).
Recent advances in fabrication technology resulted in TMRs as high
as 90\% at low temperatures and above 50\% at room temperature
\cite{tsu,fau}. However, the physics that determines the
polarization of tunneling electrons is still poorly understood.
For example, although the low voltage ($<$ 0.1V) TMR has been
steadily increased over the last few years, the effect drops to
zero around 1V and can even change sign. This voltage dependence
has been ascribed to magnon excitations \cite{sspp1,magnon}, the
detailed interface electronic structure \cite{ng,sharma,jm},
metal-induced gap states \cite{mav}, as well as the general
behavior of spin polarized free electron models for tunnel
magnetoresistance \cite{davis,mont}.

In this paper we show that the voltage dependence of MTJs can be
separated into cathode and anode effects by measuring the dc
voltage dependence of the signal in a
CoFe-Al$_2$O$_3$-Al-Al$_2$O$_3$-NiFe mesoscopic spin valve. We
found that the tunneling spin polarization \emph{from} the
ferromagnet (FM acting as cathode) is weakly voltage-dependent,
but drops strongly and even inverts with voltage when tunneling
\emph{into} it (FM acting as anode). A bias-dependent free
electron model shows that electrons tunneling from the FM
originate below the Fermi level (large polarization), whereas
electrons tunneling into the FM face hot electron states with
decreasing polarization. At even higher FM bias, spin polarization
can change sign due to wave-vector matching effects in the
transmission probability.

\begin{figure}[t]
\includegraphics[width=3.4in]{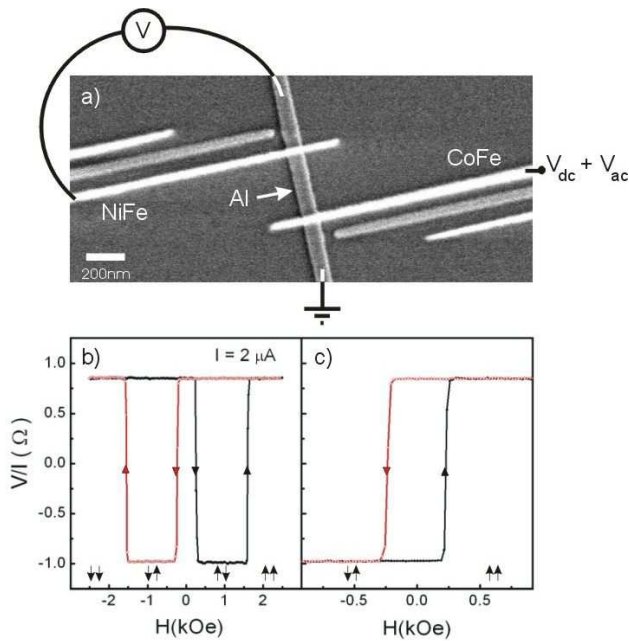}
\vspace{-12mm} \caption{a) Scanning electron microscope image of a
device. Voltage ($V_{dc}+V_{ac}$) is biased between CoFe and the
Al strip so that a current $I$ is injected from or into CoFe. The
ac voltage is measured between NiFe and the top part of the Al
strip. b) and c) Spin valve effect. Output signal $V/I$ for
$V_{dc}=0$ and $I=2 \mu$A as a function of magnetic field at 4.2
K. The arrows at the bottom of the figure represent the
configuration of the magnetization of the ferromagnetic leads.}
\label{fig1}
\end{figure}

We employ mesoscopic spin valves which have been recently used to
demonstrate electrical detection of spin precession and fixed bias
spin polarization \cite{wees}. Our spin valves (Fig. \ref{fig1}a)
consist of a CoFe ferromagnetic electrode which ``sources"
spin-polarized electrons into an aluminum (Al) strip through a
tunnel barrier. At a distance $d$ from the source there is a
second electrode (NiFe) which detects spin polarized electrons in
the Al strip by sensing a voltage (the detector is never biased in
this experiment).

A current $I$ from or into the CoFe source results in an unequal
density of spin-up and spin-down electrons in the aluminum
\cite{wees,vson} with a difference proportional to $I$ and the
effective polarization of the CoFe electrode, $P_{CoFe}$, which
characterize the difference between the majority and minority spin
populations of the electrons that participate in the tunneling (by
definition, $P_{CoFe}=\frac{I_{maj}-I_{min}}{I}$ where
$I_{maj,min}$ are the tunneling currents for majority and minority
electrons and $I=I_{maj}+I_{min}$ is the total current). This spin
imbalance will diffuse in both directions along the Al strip and
will reach the detector electrode which will sense a weighted
average of the two spin densities by its own polarization,
$P_{NiFe}$. Therefore, the output voltage $V$ is related to the
spin degree of freedom and is proportional to
$P_{CoFe}*P_{NiFe}*I$ \cite{wees,vson}. $V$ changes sign when the
FM relative magnetizations switch from parallel to antiparallel
configuration. As described below, by applying a small ac voltage
superimposed on a dc voltage and measuring the difference in the
detector ac voltage between these two configurations \cite{wees},
we analyze the dc-bias dependence of the polarization of injected
electrons from the CoFe into the Al strip or vice versa
\cite{com1}. This is analogous to studying the voltage dependence
of the TMR in MTJs separating the cathode and anode effects.

We prepare the devices with electron beam lithography and a
three-angle shadow evaporation technique to produce tunnel
barriers \emph{in situ} \cite{SOV}. An aluminum strip (100 nm wide
and 6 nm thick) is first deposited through a suspended mask onto a
Si/SiO$_2$ substrate using e-beam evaporation with a base pressure
lower than $10^{-8}$ Torr. Next, the aluminum is oxidized in pure
oxygen (150 mTorr for 20 min) to generate the insulating
Al$_{2}$O$_{3}$ barriers. Then, without breaking vacuum, the FM
electrodes (60 nm wide) are deposited sequentially from two
different angles forming tunnel junctions where they overlap with
the Al strip. The thickness of the NiFe and CoFe electrodes were
chosen to be 20 and 35 nm, respectively to geometrically enhance
the difference in the coercive fields which naturally occurs in
these materials. The resulting tunnel resistance of the junctions
is around 50 k$\Omega$. A small magnetic field of 60 Oe along the
FM electrodes was applied during the growth to favor the alignment
of their magnetocrystalline and shape anisotropies. The distance
$d$ between the FM electrodes was varied from 150 to 1,000 nm. The
data shown in this paper were acquired for $d$ = 220 nm.

The measurements were performed using lock-in techniques by
applying a voltage bias ($V_{dc}+V_{ac}$) to the source junction
(CoFe) and measuring ac voltage at the remote detector junction as
sketched in Fig. \ref{fig1}a. Fig. \ref{fig1}b shows a typical
spin-valve signal $V/I$ at 4.2 K as a function of an applied
in-plane field along the axis of the FM electrodes ($V_{dc}=0$).
At large enough negative magnetic fields the magnetization of the
FM electrodes are pointing to the same direction (parallel
configuration). As the magnetic field is swept from negative to
positive, a change in the sign of the detector signal is observed
at 250 Oe when the magnetization of the NiFe electrode reverses
and the device switches from parallel to antiparallel
configuration. As the magnetic field is further increased the CoFe
flips (at 1.5 kOe ) and a parallel configuration is recovered.
Fig. \ref{fig1}c demonstrates that both configurations (parallel
and antiparallel) are possible at zero magnetic field and that
they can be prepared in a controlled way.

\begin{figure}[t]
\vspace{17mm}
\includegraphics[width=3.8in]{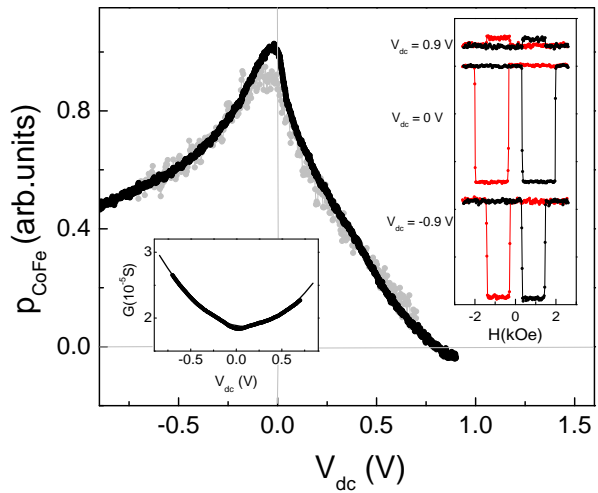}
\vspace{-25mm} \caption{Normalized dynamic polarization $p_{CoFe}$
for a sample at 4.2 K (black) and a sample at 295 K (gray).
Electrons are injected from the CoFe for $V_{dc} < 0$ and into it
for $V_{dc} > 0$. Top inset: typical spin valve effect for
different dc bias showing a change in sign in the voltage
switching at large positive bias. Bottom inset: conductance of the
CoFe junction. $V_{ac}=30$ mV.} \label{fig2}
\end{figure}

By measuring the output voltage difference between parallel and
antiparallel configurations as a function of $d$, we estimate
\cite{wees} the (low bias) polarization of the electrodes to be of
the order of 25\% at 4.2 K. Note that the larger polarization and
the reduction of sample dimensions by an order of magnitude, in
particular the distance between the FM electrodes and the Al
thickness, help increase the detection signal by a factor of 200
as compared to reported values by Jedema \emph{et al.}
\cite{wees}. This allows us to perform sensitive dc voltage
dependent measurements. In this case, the source junction is
excited with both dc and ac voltages. The small ac voltage (30 mV)
is used to sense the variation of the polarization as the dc
voltage is swept from negative to positive values. The difference
in the output voltage for the two relative configurations of the
FM electrodes with the ac technique is proportional to
$\frac{d(P_{CoFe}I)}{dV_{dc}}=\frac{d(I_{maj}-I_{min})}{dV_{dc}}=G_{maj}-G_{min}$
where $G_{maj,min}$(which are $V_{dc}$ dependent) are the
\emph{dynamic} conductances of the majority and minority electrons
and thus $G_{maj}+G_{min}$ is the total dynamic conductance of the
CoFe junction. Changes in the transmission of majority and
minority spin electrons as a function of bias are clearly
illustrated by the \emph{dynamic} polarization defined as
$p_{CoFe} = (G_{maj}-G_{min})/(G_{maj}+G_{min})$ \cite{com2}.
$p_{CoFe}$ can then be obtained by dividing the difference in the
output voltage for the two relative configurations of the FM
electrodes by the dynamic conductance of the CoFe junction (bottom
inset in Fig. \ref{fig2}). The magnetic configuration is prepared
as shown in Fig. \ref{fig1}c.

The main panel of Fig. \ref{fig2} shows $p_{CoFe}$ for two
different samples: one measured at 4.2 K and the other at room
temperature. The same bias dependence of $p_{CoFe}$ is observed
for both samples and for 4 other samples not presented here. For
negative dc biases, electrons are injected from the CoFe electrode
into the Al source region. The resulting signal, shown in the main
panel of Fig. \ref{fig2}, reaches a maximum around -50 mV and then
drops but, even for the largest negative voltages that were
applied, the signal is still of comparable magnitude to the one at
zero bias. On the other hand, when the electrons are injected from
the Al into the CoFe electrode (positive bias), the dynamic
polarization drops faster. When $V_{dc}=+0.5$ V, the detector
signal has decreased significantly; it reaches zero at
approximately +0.8 V, and it is clearly negative at +0.9 V. The
top inset of Fig. \ref{fig2} shows the output voltage as the
magnetic field is swept for different values of the voltage bias.
At $V_{dc}=+0.9$ V, a change in the sign in the voltage switching
is clearly seen which implies that, around that bias, the dynamic
conductance for minority electrons dominates.

Using a free electron model described below, we calculate the spin
currents in the CoFe-Al$_{2}$O$_{3}$-Al source junction versus
bias, and discuss how these results are related to the experiments
above and to the magnetoresistance in magnetic tunnel junctions in
general.

\begin{figure}[t]
\includegraphics[width=3.7in]{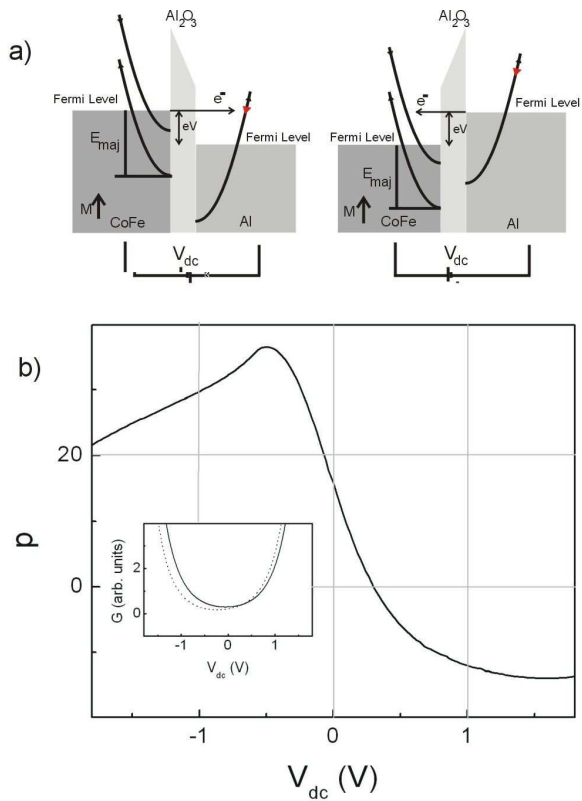}
\vspace{-5mm} \caption{a) Schematic diagram of the source tunnel
junction. Parabolic $E(k)$ curves displaced in energy represent
the exchange-split majority and minority spin bands of the
ferromagnetic electrode, CoFe. Left (right) panel: representation
of negative (positive) CoFe bias. b) Calculated dynamic
polarization $p$ as a function of bias. The inset shows the
dynamic conductance of the junction for the majority (solid line)
and minority band (dotted line). } \label{fig3}
\end{figure}

The tunnel current through a barrier at finite bias
$j(V)=j_{LR}-j_{RL}$ is given by \cite{duk},

\begin{eqnarray}
j=\frac{2e}{h}\int
dE[f_{L}(E)-f_{R}(E+eV)]\int\int\frac{d^{2}k_{\parallel}}{(2\pi)^{2}}D(eV,E,k_{\parallel})\nonumber
\end{eqnarray}
\\
where $j_{LR}$ ($j_{RL}$) are the currents from the left (right)
to right (left) electrodes, $D(eV, E, k_{\parallel})$ is the
bias-dependent transmission probability, $f(E)$ the usual
Fermi-Dirac function, $e$ the electron charge and $h$ Planck's
constant. $D(eV, E, k_{\parallel})$ is found by exactly solving
the Schr\"{o}dinger equation for parabolic free-electron energy
bands and a trapezoidal shape tunnel barrier using the Airy
wave-function solutions \cite{gun} within the tunnel barrier (Fig.
\ref{fig3}a). In order to calculate the current, first $D$ is
computed by integrating over all electrons with the same energy,
$E$ but varying $k_{\parallel}$, the component of $k$ parallel to
the barrier. Then $j(V)$ is found by integrating over $E$. As
shown in Fig. \ref{fig3}a, the ferromagnetic electrode is
characterized by an exchange-split parabolic energy band, which
results in majority and minority bands with distinct energies
$E_{maj}$ and $E_{min}$ for the bottom of the bands below the
Fermi energy. Note that for Al$_2$O$_3$ barriers, it is likely
that the electron bands largely responsible for the tunneling
current will have significant $s-p$ character and thus a
free-electron like dispersion relation ($E\propto k^2$) is
justified \cite{davis,st}.

The offset of the FM energy bands responsible for tunneling have
been derived for Fe from first principles band structure
calculations, namely, $E_{maj}\sim2.25$ eV and $E_{min} \sim 0.5$
eV \cite{davis,st}. For our calculations we use these values, a
Fermi energy for the Al, $E_{F,unp}=11.7$ eV, and a barrier with a
thickness of 1.2 nm and a height $\phi = 1.35$ eV \cite{davis}.
Since spin-flip scattering is neglected, the total current is
comprised of two independent electron-tunneling currents
associated with spin-up and spin-down electrons. For the
negatively biased CoFe-Al$_2$O$_3$-Al junction, the total current
is $I=I_{maj\rightarrow unp}+I_{min\rightarrow unp}$, where
$I_{maj(min)\rightarrow unp}$ is the current from the majority
(minority) band to the unpolarized band in the normal metal (Al in
our experiment). To compare the simulations with the experimental
results in Fig. \ref{fig2}, we calculate the dynamic conductances
$G_{maj(min)} = \frac{dJ_{maj(min)\rightarrow unp}}{dV}\; (V <0);
\frac{dJ_{unp\rightarrow maj(min)}}{dV}\; (V >0)$ and the dynamic
polarization as defined above $p =
(G_{maj}-G_{min})/(G_{maj}+G_{min})$.

As seen in Fig. \ref{fig3}b, the calculated dynamic polarization
qualitatively follows all the essential features of the
measurements in Fig. \ref{fig2} allowing us to interpret their
physical origin. The bias dependence at large negative voltage is
weak as in the experimental results; this is a consequence of the
narrow energy distribution of the injected electrons around the
Fermi energy in the FM. More interestingly, calculations at
positive bias show a large bias dependence of the dynamic
polarization as well as a sign change as found experimentally. By
analyzing the tunneling for the majority and minority spin band
electrons, we conclude that the large drop in $p$ is partially
caused by the decrease of density of states ratio between the two
spin bands for electrons tunneling into hot states, as shown in
the right panel of Fig. \ref{fig3}a. However, it is evident from
inset in Fig. \ref{fig3}b that $G_{min}$ becomes larger than
$G_{maj}$ above $\sim$0.4 V, causing a sign change. A sign
inversion cannot be explained with only a decreasing density of
states ratio between minority and majority band electrons.
Instead, we have found numerically that the transmission through
the barrier is largest when the total electron energy is equal to
the effective barrier height (this is when the absolute value of
electron and (imaginary) barrier wave vectors are equal), creating
a wave vector matched state which will occur at different biases
for the two spin bands in the FM. Specifically, around 0.5 V,
$E_{min}+eV = (0.5+0.5)$ eV = 1 eV is closer to $\phi=1.35$ eV
than $E_{maj}+eV = (2.5+0.5)$ eV = 3 eV. In other words, the
matching of the minority band improves with increasing bias,
whereas the matching of the majority band deteriorates. As a
consequence the relative increase in the tunneling of minority
electrons is faster than expected by simple density of state
considerations.

Note that the measurements in Fig. \ref{fig2} seem displaced
towards positive biases as compared to the simulations in Fig.
\ref{fig3}b. This could be accounted by an asymmetry between the
CoFe-Al$_2$O$_3$, Al$_2$O$_3$-Al barrier heights that is not
considered in our model \cite{brinkman}.

We find that our experimental and numerical separation of the
ferromagnetic electrodes forms a solid basis for understanding the
physical processes responsible for voltage dependence of TMR and
tunnel spin injection-accumulation in general. The presented
results demonstrate that even when TMR is zero, the transfer of
spin polarized electrons from the cathode is still possible and
that the anode is responsible for the quenching of the
magnetoresistance. This could explain recent results by Fuchs
\emph{et al.} \cite{fuc} in nanoscale MTJs where it was shown that
spin-transfer switching can occur in the absence of TMR at high
bias, meaning that the junction current is still polarized. It is
also worth noting that interesting and complex bias dependencies
can be obtained by engineering magnetic tunnel barriers interfaces
and using composite barriers \cite{fau,sharma,jm}. This
illustrates the importance of studying all the aspects of the
quantum mechanical properties of the tunneling process before
ascribing complex results to an unusual density of states of the
material.

In conclusion, we have measured the voltage dependence of the
tunneling polarization of electrons from and into a ferromagnet
emphasizing the intrinsic asymmetry between these two processes.
We found that the polarization is strongly suppressed for
electrons tunneling into the ferromagnet due to the reduced
polarization for hot electron states and a spin-dependent
wave-vector mismatch.

This research was supported in part by NSF grants DMR-0244441 and
NSEC-PHY-0117795 and ONR grant N00014-02-1-0055.


\begin{thebibliography}
\bibitem{com1} Note that the roles of the FM electrodes are interchangeable. The
NiFe and CoFe electrodes can both be used as source or detector.
In either case, the results are similar; we focus on the tunneling
through the CoFe.
\bibitem{com2} Note that the dynamic polarization, $p_{CoFe}$, measures
contributions from electrons within a narrow energy range for each
bias voltage whereas the current polarization, $P_{CoFe}$,
includes contributions from all electrons.
\end{thebibliography}
\end{document}